\documentclass[english,aps,prX,showpacs,twocolumn,superscriptaddress]{revtex4-2}
\usepackage[T1]{fontenc}
\usepackage[utf8]{inputenc}
\usepackage{babel}
\usepackage{graphicx}
\usepackage{bm}
\usepackage{bbm}
\usepackage{array}
\usepackage{amsmath}
\usepackage{amssymb}
\usepackage{dcolumn}
\usepackage{epstopdf}
\usepackage{bbold}
\usepackage{tikz}
\usepackage{pgfplots}
\usepackage[margin=15mm]{geometry}
\colorlet{linkequation}{blue}
\usepackage[colorlinks=true,linkcolor=blue,citecolor=blue,urlcolor=blue]{hyperref}
\usetikzlibrary{shadings}
\usepackage{siunitx}
\usepackage[nohyperlinks, nolist]{acronym}
\usepackage{braket}
\usepackage{mathrsfs}
\usepackage{import}
\usepackage{filecontents}
\usepackage{color}
\usepackage{miller}

\newcommand{\kB}[0]{k_{\mathrm{B}}}

\usepackage[normalem]{ulem}

\begin{document}

\title{Anisotropic skyrmion liquid phase}

\author{Daniel Schick}
\email[]{daniel.schick@uni-konstanz.de}
\affiliation{Fachbereich Physik, Universit\"at Konstanz, DE-78457 Konstanz, Germany}

\author{Tim Matthies}
\affiliation{Department of Physics, University of Hamburg, DE-20355 Hamburg, Germany}

\author{Thomas Mutschler}
\affiliation{Fachbereich Physik, Universit\"at Konstanz, DE-78457 Konstanz, Germany}

\author{Levente R\'{o}zsa}
\affiliation{Department of Theoretical Solid State Physics, Institute of Solid State Physics and Optics, HUN-REN Wigner Research Centre for Physics, H-1525 Budapest, Hungary}
\affiliation{Department of Theoretical Physics, Budapest University of Technology and Economics, M\H{u}egyetem rakpart 3., H-1111 Budapest, Hungary}

%\author{Markus Wei{\ss}enhofer}
%\email[]{markus.weissenhofer@uni-konstanz.de}
%\affiliation{Fachbereich Physik, Universit\"at Konstanz, DE-78457 Konstanz, Germany}

\author{Ulrich Nowak}
\affiliation{Fachbereich Physik, Universit\"at Konstanz, DE-78457 Konstanz, Germany}

\date{\today}

\renewcommand{\figurename}{FIG.}

%------------------------------------------------------------------------
%TC:ignore
\begin{abstract}
The nature of the melting transition in two-dimensional systems of particles has attracted considerable research attention since the development of \ac{KTHNY} theory. The hexatic phase proposed by this theory has been recently identified experimentally in ensembles of magnetic skyrmions, quasiparticles formed in a magnetically ordered crystal. Here, we use quasiparticle dynamical simulations to study how the anisotropy of the skyrmion-skyrmion interactions induced by the atomic lattice influences the melting transition. For isotropic interactions, we find a transition from a solid phase through a hexatic phase stable in a narrow temperature range to an isotropic liquid phase. However, if the interactions between skyrmions are forced to be anisotropic by the atomic lattice, then a direct solid-liquid transition can be observed with orientational order persisting up to temperatures of 30~K in the liquid phase.
%Our system breaks continuous rotational symmetry in the liquid phase
\end{abstract}
%TC:endignore
%-------------------------------------------------------------------------

%\keywords{..., }

\maketitle
\begin{acronym}
\acro{DMI}[DMI]{Dzyaloshinsky-Moriya interaction}
\acro{sLLG}[sLLG]{stochastic Landau-Lifshitz-Gilbert}
\acro{LLG}[LLG]{Landau-Lifshitz-Gilbert}
\acro{KTHNY}[KTHNY]{Kosterlitz-Thouless-Halperin-Nelson-Young}
\end{acronym}

\section{Introduction}
Phase transitions in ensembles of particles have been studied extensively in a wide variety of physical systems. In three-dimensional materials, crystals displaying long-ranged positional and orientational order of the particles melt into isotropic liquids where only short-ranged correlations are present through a first-order transition. In two dimensions, the presence of a third phase of matter was proposed, the hexatic phase with short-range positional and long-range orientational order between the quasi-long-range-ordered solid and the isotropic liquid phases. The phase transitions between these three phases are continuous according to the \ac{KTHNY} theory~\cite{Kosterlitz_1973, Halperin_1978, Nelson_1979, Young_1979}. This theory also establishes that the unbinding of topological defects, including dislocations and disclinations, provides the microscopic mechanism for these phase transitions. It was later recognized that even if the hexatic phase is present, it may be separated from the liquid or the solid state by a first-order transition instead of the second-order one predicted by the \ac{KTHNY} theory~\cite{Bladon1994,Bernard2011}. Which of these scenarios is realized in a system depends on the microscopic details, for example the presence or absence of an underlying periodic potential~\cite{Halperin_1978,Nelson_1979}, whether the interactions between the particles is repulsive at all distances or contains a region of attraction~\cite{Bladon1994}, and how fast this interaction decays with the distance~\cite{Kapfer2015}. While most studies have focused on particles with circularly symmetric interaction potentials, various phase transitions have been identified in ensembles of non-circular hard particles in simulations~\cite{Wojciechowski2004, Anderson2017} and experiments~\cite{Walsh_2016}. The nature of these phase transitions has stimulated a considerable amount of theoretical and simulational studies, as well as experimental works where the hexatic phase has been identified in systems of electrons~\cite{Guo1983}, atoms~\cite{Negulyaev2009}, molecules forming liquid crystals~\cite{Cheng1988}, colloidal particles~\cite{Murray1987,Armstrong1989,Marcus1996,Zahn1999}, superconducting vortices~\cite{Guillamon2009,Roy2019}, and vortices in a quasi-two-dimensional Bose gas \cite{PhysRevLett.133.143401}.

Magnetic skyrmions are localized spin configurations in magnetically ordered crystals~\cite{Bogdanov1989,Bogdanov_1994,Muehlbauer2009,Yu2010}. They are exceptionally stable against perturbations, making it possible to treat them as quasiparticles with a conserved particle number. Skyrmions form a triangular lattice in the ground state, corresponding to a solid phase at finite temperature, which melts into a liquid state as the temperature is increased. Experimental studies have also identified an intermediate hexatic phase between these phases~\cite{Seshadri1992,Huang2020,Gruber2025}. However, skyrmions differ from other particles for which two-dimensional melting has been studied in several aspects. First, the skyrmions are formed on an underlying lattice of magnetic atoms. Therefore, they interact with defects in the atomic lattice~\cite{Reichhardt2022}, and they also experience a periodic potential due to this lattice. If the size of skyrmions is comparable to the atomic separation, this periodic potential can become sufficiently strong to exclude the formation of a hexatic phase, as was found in numerical simulations of the melting transition~\cite{Nishikawa2019}. Second, skyrmions are characterized by a topological charge influencing their dynamics, which also affects the velocity of defects in the skyrmion crystal which determines how fast the melting occurs~\cite{Schick2024}. Third, the size of skyrmions, thereby the strength of the interaction between them, may be adjusted by the external magnetic field, which can be utilized to drive the phase transitions~\cite{Huang2020,Gruber2025}. Fourth, the underlying microscopic interactions between the spins determines the effective interaction potential between the skyrmions. Studies of the melting transition so far have focused on skyrmions repulsing each other at all distances~\cite{Nishikawa2019,Huang2020,Gruber2025}, which are stabilized by the Dzyaloshinsky--Moriya interaction~\cite{Dzyaloshinsky1958,Moriya1960,Lin2013}. However, skyrmions stabilized by the competition between ferromagnetic and antiferromagnetic isotropic exchange interactions also exhibit attraction at certain distances~\cite{Leonov2015}, and the interaction between them may be strongly anisotropic due to the underlying atomic lattice~\cite{Rozsa2016}.

Here, we study the influence of the anisotropic interaction between the skyrmions on the melting of the skyrmion lattice. We determine the interaction potential from atomistic spin interactions describing the (Pt$_{0.95}$Ir$_{0.05}$)/Fe/Pd(111) system. We use this anisotropic interaction potential and its isotropic counterpart to perform Brownian dynamics simulations of skyrmions at temperatures up to 30~K in increments of at most 1~K. We calculate positional and orientational order parameters, as well as the numbers of dislocations and disclinations, to characterize the transitions. We find that the anisotropy induced by the underlying atomic lattice acts as an agent promoting positional and orientational order, similarly to how magnetic or electric fields may be used to align magnetic or electric dipoles, or alignment layers at the surfaces of liquid crystals stabilize the orientation of the molecules. Consequently, the anisotropic interactions stabilize the solid phase up to higher temperature than the isotropic interactions, while orientational order remains observable up to the highest simulated temperatures in the liquid phase despite the presence of a large number of topological lattice defects. %We find that the anisotropic interactions enhance the stability of the ordered phases. Furthermore, we find that %the system does not become completely isotropic even in the liquid phase
%anisotropic bond orientations remain observable even in the presence of a large number of topological lattice defects due to the anisotropy of the interactions being locked to the crystallographic axes of the underlying atomic lattice, resulting in an anisotropic liquid-like behavior. 
We compare the phases determined from the correlation functions to a machine-learning classification of the observed spin configurations.
%Magnetic skyrmions are localized topological spin-structures \cite{Sondhi_1993, Bogdanov_1994, Muehlbauer2009}, which behave as quasiparticles at finite temperatures. They are candidatates for a multitude of novel technologies in data storage and computing \cite{Pinna_2018, Zazvorka_2019}. They can also serve as an ideal candidate for investigating two-dimensional processes, such as phase transitions which are described in the \ac{KTHNY}-theory~\cite{Kosterlitz_1973, Halperin_1978, Nelson_1979, Young_1979}.

\section{Methods}

\subsection{Atomistic spin model\label{sec:2a}}

We consider the (Pt$_{0.95}$Ir$_{0.05}$)/Fe/Pd(111) system in our simulations. The magnetic Fe atoms are described by the atomistic spin Hamiltonian
\begin{align}
    \mathcal{H} = \frac{1}{2} \sum_{i \neq j} \bm{S}_i \mathcal{J}_{ij} \bm{S}_j + \sum_i \bm{S}_i \mathcal{K} \bm{S}_i - \mu_S \sum_i \bm{S}_i \cdot \bm{B},
\end{align}
with the exchange interaction tensors $\mathcal{J}_{ij}$, the anisotropy tensor $\mathcal{K}$, external magnetic field $\bm{B}$ and magnetic spin moment $\mu_S$. The material parameters were determined from first-principles calculations in Ref.~\cite{Rozsa2016}, and are also reported in Ref.~\cite{Zazvorka_2019}. The lattice constant of the triangular lattice formed by the Fe atoms is $a=2.751$~\AA. We apply an out-of-plane magnetic field of $B=1$~T, where skyrmions exist as metastable quasiparticles in a field-polarized ground state. We determine the interaction potential by inserting two skyrmions in the collinear background, fixing the centers of the skyrmions at certain lattice sites by aligning the spins at those sites opposite to the external field, and relax all other spin directions at zero temperature using atomistic spin-dynamics simulations until a local energy minimum is reached. We set the zero of the potential energy at the point where the skyrmions are sufficiently far away that the interaction becomes negligible.

\subsection{Dynamical simulations\label{sec:2b}}

Neglecting the internal degrees of freedom of skyrmions in a rigid-body approximation, their dynamics as quasiparticles is governed by the Thiele equation~\cite{Thiele1973},
\begin{align}
    \bm{\mathcal{G}} \times \bm{v}_i +\alpha\mathcal{D}\bm{v}_i = \bm{F}_i,
\end{align}
where $\bm{\mathcal{G}} = \mathcal{G} \bm{e}_\perp$ is the gyrocoupling vector, $\mathcal{D}$ is the dissipation coefficient, and $\alpha$ is the Gilbert damping known from the Landau-Lifshitz-Gilbert equation. We use the parameters $\mathcal{G}/k_{\textrm{B}} = 2.272 \ \mathrm{ns \ K \ nm^{-2}}$ and $\mathcal{D}/k_{\textrm{B}} = 3.396\ \mathrm{ns \ K \ nm^{-2}}$ determined in Ref.~\cite{Weissenhofer2020}, and $\alpha=0.1$. On the right-hand side of the equation, the force $\bm{F}_i$ acting on the skyrmion comprises of a term from the skyrmion-skyrmion interaction $\bm{F}_{i,\mathrm{sk-sk}}$ and thermal fluctuations $\bm{F}_{i,\mathrm{th}}$. %, i.e.
% \begin{align*}
%     \bm{F}_{i} = \bm{F}_{i,\mathrm{sk-sk}} + \bm{F}_{i,\mathrm{th}}
% \end{align*}
The thermal force is modeled as Gaussian white noise with expectation value $\langle \bm{F}_i(t) \rangle = 0$ and covariance $\langle F_i^{\mu}(t) F_j^{\nu}(t^\prime) \rangle = 2 \alpha \kB T \mathcal{D}_{\mu \nu} \delta_{ij} \delta(t- t^\prime) $. The quasiparticle description is assumed to hold within the considered temperature range. For sufficiently high temperatures (above 50~K for the considered material~\cite{PhysRevB.103.214417}), thermal fluctuations are expected to destabilize the skyrmion configuration, invalidating the quasiparticle approximation on the time scale of the simulations performed here. \\
The equation of motion is solved numerically for 16680 interacting skyrmions in a box size of $272.9~\mathrm{nm} \times 273.4~\mathrm{nm}$ with periodic boundary conditions. The system is thermalized for $25.37~\mathrm{ns}$ before the equilibrium properties are extracted by averaging over $120.8~\mathrm{ns}$.

\subsection{Description of phase transitions\label{sec:2c}}

The \ac{KTHNY} theory describes the melting process through the unbinding of topological lattice defects. These are identified in our simulations by counting the number of neighbors of each skyrmion in a given configuration via Voronoi tessellation. In a close-packed configuration, all skyrmions have six neighbors. %Slightly displacing one skyrmion 
Moving two next-nearest-neighboring skyrmions closer to each other results in two skyrmions with seven neighbors and two skyrmions with five neighbors forming a topologically trivial defect known as a dislocation pair. These defects may split into pairs of skyrmions with five and seven neighbors known as dislocations. Dislocations are characterized by the Burgers' vector and are topologically non-trivial; they may only vanish if two dislocations with opposite Burgers' vector merge. A single skyrmion with a number of neighbors different from six, where each of its neighbors has six neighbors, is known as a disclination. 

Identifying the neighbors of each skyrmion makes it possible to calculate the orientational bond-order parameter
\begin{align}
    \Psi_{6}(\bm{r}_i) = \frac{1}{n_i} \sum_j e^{-6 \mathrm{i} \theta_{ij}},
\end{align}
which is calculated for skyrmion $i$ by an average over all of its neighbors $j$, where the angle $\theta_{ij}$ is defined between the relative position $\bm{r}_j - \bm{r}_i$ and a fixed reference direction in the plane \cite{Bernard2011}. The spatial and temporal orientational correlation function is defined as \cite{PhysRevE.75.031402, PhysRevE.73.066106}
\begin{align}
    g_6(|\bm{r}_i - \bm{r}_j|) &= \langle \Psi_{6}^*(\bm{r}_i) \Psi_{6}(\bm{r}_j) \rangle, \label{eq:g6r} \\
    g_6(t_2-t_1) &= \langle \Psi^*(t_2) \Psi(t_1) \rangle, \label{eq:g6t}
\end{align}
while the positional order can be characterized by the pair correlation function \cite{PhysRevE.73.066106}
% \begin{align*}
%     g(|\bm{r}_1-\bm{r}_2|) = \frac{\langle \rho (|\bm{r}_1-\bm{r}_2|) \rangle}{\rho_0}. \label{eq:pairCorrelation}
% \end{align*}
\begin{align}
    g(r) = \frac{A}{4 \pi r N^2} \langle \sum_{i} \sum_{j\neq i} \delta(r - r_{ij}) \rangle, \label{eq:pairCorrelation}
\end{align}
    where $A$ is the system area, $N$ the number of skyrmions, and $r_{ij}$ the distance between skyrmions $i$ and $j$. We further characterize the positional correlations by calculating the structure factor
    \begin{align}
    S(\bm{k}) = \frac{1}{N} \left\langle \left| \sum_{i,j} e^{- \textrm{i}\bm{k} \cdot (\bm{r}_i-\bm{r}_j)} \right| \right\rangle \label{eq:structure_factor}
\end{align}
in reciprocal space.

In the %quasi-crystalline 
solid phase, one finds quasi-long-ranged positional order, with the pair correlation showing algebraic decay ($g(r) \propto r^{-\nu}$), whereas the orientational correlation reaches a finite value at $r \rightarrow \infty$. This implies the existence of a global order and thus a finite value for the average bond order $|\langle \Psi_6 \rangle|$, which vanishes in phases with finite orientational correlation length. In the hexatic phase, positional order is short-ranged with the correlations decaying exponentially, while the long-ranged orientational correlations follow an algebraic decay. The isotropic liquid phase is characterized by short-ranged positional and orientational correlations.

Within \ac{KTHNY} theory, the transition from the solid to the hexatic phase is accompanied by the unbinding of dislocation pairs into dislocations, while the hexatic to isotropic liquid transition occurs where the dislocations unbind into disclinations. However, the phase transition determined from the correlation functions may also occur at different parameter values than the onset of a certain type of defects, particularly if the phase transition is of first order contrary to \ac{KTHNY} theory; see, e.g., Ref.~\cite{Anderson2017} for anisotropic hard particles. Since here we study the effect of the anisotropy of the skyrmion-skyrmion interactions on the melting, we analyze the correlation functions and the number of topological defects separately, and compare their behavior with increasing temperature.

\section{Results}
\subsection{Skyrmion-skyrmion interaction\label{sec:3a}}
\begin{figure}
    \centering
    \includegraphics[width=0.9\linewidth]{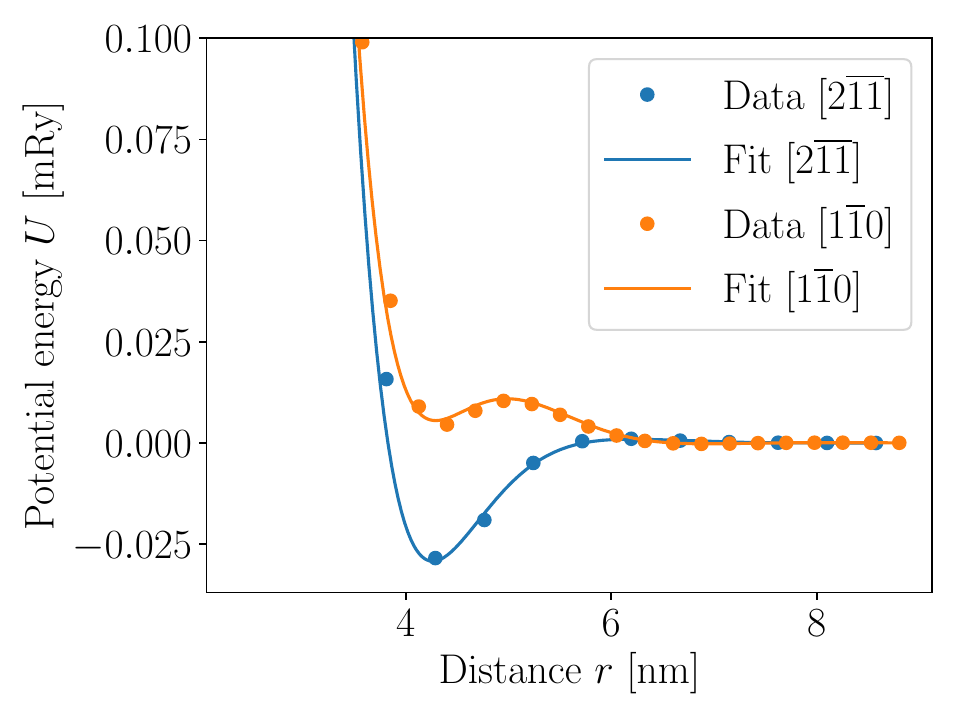}
    \caption{Interaction potential between two skyrmions in the (Pt$_{0.95}$Ir$_{0.05}$)/Fe/Pd(111) system at $B= 1 \ \mathrm{T}$ along the $[2\overline{1}\overline{1}]$ and $[1\overline{1}0]$ directions. Dots denote simulation data points, whereas lines indicate fits according to Eq.~\eqref{eq:PotentialFit}.}
    \label{fig:Potential_Fit}
\end{figure}

% We determine our interaction potential between skyrmions by atomistic spin simulation. The corresponding Hamiltonian in the Fe layer reads as
% \begin{align}
%     \mathcal{H} = \frac{1}{2} \sum_{i \neq j} \bm{S}_i \mathcal{J}_{ij} \bm{S}_j + \sum_i \bm{S}_i \mathcal{K} \bm{S}_i - \mu_S \sum_i \bm{S}_i \cdot \bm{B},
% \end{align}
% with the coupling tensors $\mathcal{J}_{ij}$, the anisotropy tensor $\mathcal{K}$, external magnetic field $\bm{B}$ and magnetic spin moment $\mu_S$. The corresponding parameters are taken from \cite{Rozsa2016}. Using atomistic spin simulations, we can insert two skyrmions into a colinear background, fixing the center of the skyrmions, and letting them relax at zero temperature. 
%From this, we can subtract the colinear state energy and receive an interaction potential, which can be seen in FIG. \ref{fig:Potential_Fit}, alongside a fit, which we use to model our potential. This fit is of the form
We calculate the interaction potential between pairs of skyrmions in the (Pt$_{0.95}$Ir$_{0.05}$)/Fe/Pd(111) system as described in Sec.~\ref{sec:2a}. We position the skyrmions along the nearest-neighbor $[1\overline{1}0]$ and the next-nearest-neighbor $[2\overline{1}\overline{1}]$ directions on the atomic lattice, so that the restriction that the centers of the skyrmions are fixed at lattice sites results in the densest distribution of points. The obtained interaction potentials are shown in Fig.~\ref{fig:Potential_Fit}. We fit the potential curves to the model function
\begin{align}
    U(r_i) = e^{- \kappa_i r_i}(A_i \cos(k_i r_i + \phi_i) + B_i r_i + C_i), \label{eq:PotentialFit}
\end{align}
    where $r_i$ is the absolute distance between the skyrmions along the direction $i\in\left\{[1\overline{1}0],[2\overline{1}\overline{1}]\right\}$. We show the obtained fitting values in Table \ref{tab:fit_vals}. Since the skyrmions are only about 10 lattice constants in diameter, the interaction between them is strongly affected by the geometry of the underlying atomic lattice. The isotropic exchange interactions between the atomic spins along the $[1\overline{1}0]$ and $[2\overline{1}\overline{1}]$ are ferromagnetic and antiferromagnetic, respectively~\cite{Rozsa2016}. %and therefore are distorted easily by the geometry of the lattice. 
    Therefore, we get different potential curves along the $[2\overline{1}\overline{1}]$ and $[1\overline{1}0]$ directions, similarly to the (Pt$_{0.90}$Ir$_{0.10}$)/Fe/Pd(111) system with slightly different composition in Ref.~\cite{Rozsa2016}. %While in the \hkl[2-1-1] direction, the potential is repulsive at short distances and attractive at longer ones, in the \hkl[2-1-1] direction, we get a more complicated behavior, with repulsive at short, attractive at intermediate and repulsive again at long ranges. 
    \begin{table}[htb]
    \begin{tabular}{|c|c|c|} \hline  
     \rule{0pt}{2.5ex} Parameter & value $i = [2\overline{1}\overline{1}]$ & value $i = [1\overline{1}0]$ \\ [1pt] \hline
     $\kappa_i$& $1.8491 \, \mathrm{nm}^{-1}$ & $1.5449 \, \mathrm{nm}^{-1}$\\ \hline
     $A_i$ & $-128.3853 \, \mathrm{mRy}$ & $19.9772 \, \mathrm{mRy}$ \\ \hline
     $k_i$ & $1.5525 \, \mathrm{nm}^{-1}$  & $2.1950 \, \mathrm{nm}^{-1}$ \\  \hline
     $\phi_i$& $-1.1404$ & $0.1404$ \\  \hline
     $B_i$ & $0.6074 \, \mathrm{mRy} \,\mathrm{nm}^{-1}$  &  $-5.3562 \, \mathrm{mRy} \,\mathrm{nm}^{-1}$ \\  \hline
     $C_i$ & $8.9024 \, \mathrm{mRy}$ & $46.4040 \, \mathrm{mRy}$\\  \hline
    \end{tabular}
    \caption{Interaction potential fit parameters for (Pt$_{0.95}$Ir$_{0.05}$)/Fe/Pd(111) according to Eq.~\eqref{eq:PotentialFit}. Graphical representation can be seen in Fig.~\ref{fig:Potential_Fit}.}
    \label{tab:fit_vals}
    \end{table}
    The interaction potential asymptotically follows an exponential decay modulated by an oscillating function due to the competition between the ferromagnetic and antiferromagnetic isotropic exchange interactions in the system~\cite{Leonov2015}. This results in alternating repulsive and attractive parts of the potential, with the first potential minimum being much more shallow along the $[1\overline{1}0]$ than the $[2\overline{1}\overline{1}]$ direction. At the particle densities where the melting transition can be observed, the skyrmions are expected to explore only a small part of the interaction potential~\cite{Kapfer2015}, meaning that these first potential minima are likely to play a decisive role in determining the nature of the transition. To generate a potential in two-dimensional space required for the quasiparticle dynamical simulations, %for any distance vector, 
    we interpolate between the curves along the $[1\overline{1}0]$ and $[2\overline{1}\overline{1}]$ directions in the in-plane polar angle $\varphi$ as %using squared trigonometric functions:
    \begin{align}
        U(\bm{r}) &= \cos^2(3\varphi) U_{[1\overline{1}0]}(r) +  \sin^2(3\varphi) U_{[2\overline{1}\overline{1}]}(r), \label{eq:interpolation} \\
        \text{with} \quad\bm{r} &= \left( \begin{matrix} r \cos  \varphi \\ r \sin  \varphi \end{matrix} \right).
    \end{align}

\begin{figure}
    \centering
    \includegraphics[width=0.9\linewidth]{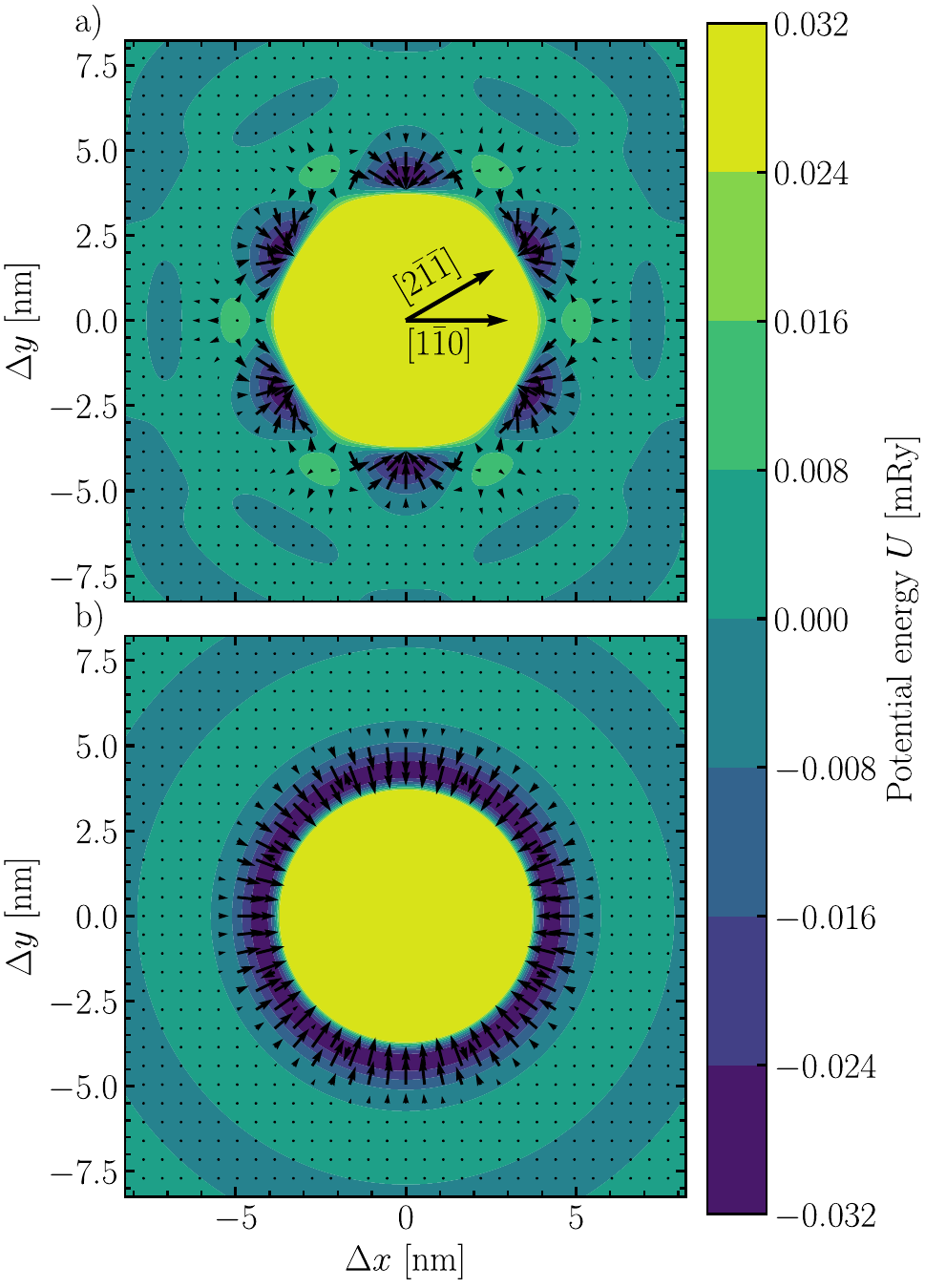}
    \caption{Interaction potentials between two skyrmions in the (Pt$_{0.95}$Ir$_{0.05}$)/Fe/Pd(111) system at $B= 1 \ \mathrm{T}$. (a) Anisotropic interaction potential obtained by interpolating between the fits along the high-symmetry directions in Fig.~\ref{fig:Potential_Fit} using the function in Eq.~\eqref{eq:interpolation}. %we use the fits from Figure \ref{fig:Potential_Fit} with the interpolation from eq. \eqref{eq:interpolation}, to get a realistic, anistropic interaction potential. In b), we establish an isotropic test case using the \hkl[2-1-1] direction from Figure \ref{fig:Potential_Fit}.
    (b) Isotropic potential obtained from the fit along the $[2\overline{1}\overline{1}]$ direction in Fig.~\ref{fig:Potential_Fit}. Arrows indicate the forces in the potential landscape.}
    \label{fig:Potentials}
\end{figure}

    %With this equation, we receive the potential described 
    The interpolated potential is shown in Fig.~\ref{fig:Potentials}(a). We see that the interaction potential has a sixfold symmetry reflecting the underlying triangular lattice, and that the skyrmions prefer to form bonds along the $[2\overline{1}\overline{1}]$ or symmetrically equivalent directions. %remain in a certain direction relative to each other. 
    At small distances $r$, we replace the interpolated potential \eqref{eq:interpolation} with a high constant value. The skyrmions are unable to explore this part of the interaction potential during the quasiparticle simulations, since it is separated by an extremely high energy barrier from the region at higher separation of the skyrmions. %and not relevant to any simulation, as the energy barrier is extremly large, such that in none of our simulations, such short distances should be achieved. \\ 
    For comparison with this anisotropic interaction potential, we also construct an isotropic potential by rotating the potential curve obtained along the $[2\overline{1}\overline{1}]$ direction around the out-of-plane direction, as illustrated in Fig.~\ref{fig:Potentials}(b). Since the skyrmions preferably form bonds along the same $[2\overline{1}\overline{1}]$ direction in the anisotropic potential, the preferred distance between the skyrmions remains the same between the two potentials along with the radial restoring forces, enabling us to primarily investigate the role of the anisotropy of the potential on the melting process. %, we present another potential, where we take the \hkl[2-1-1] part of the interaction, but apply it to all angles, so that we receive an isotropic potential. We use this as a test case against our simulation based, anisotropic interaction potential. \\

%    \section{Skyrmion lattice simulations}
\subsection{Simulation of the melting transition}

    \begin{figure}
    \centering
    \includegraphics[width=0.9\linewidth]{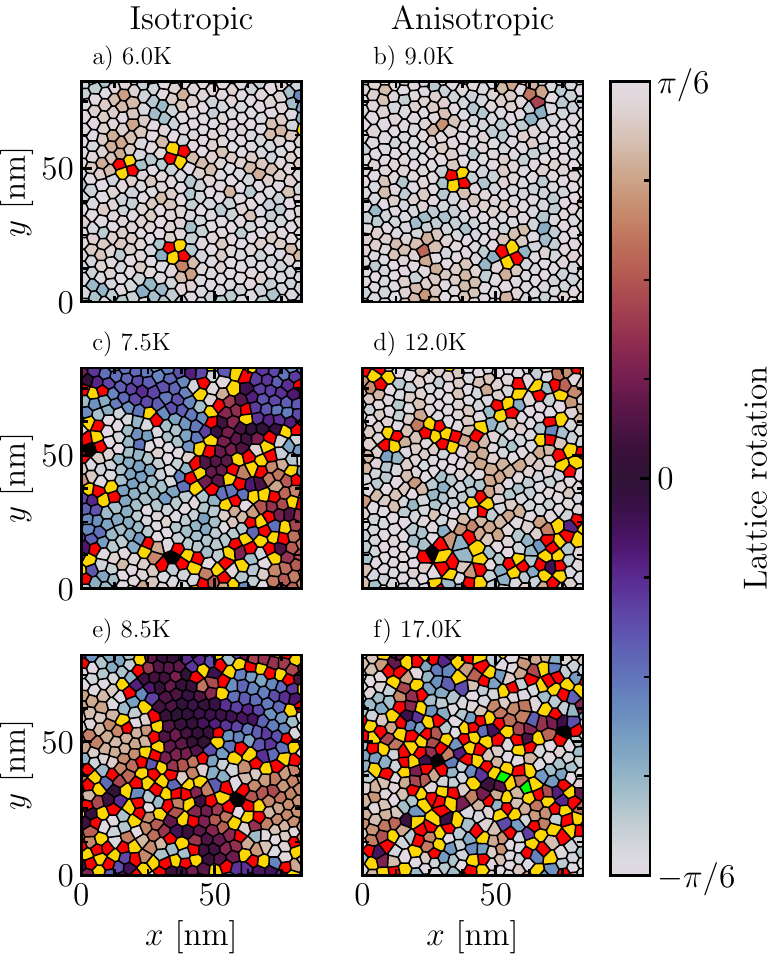}
    %\caption{Examples of simulations of skyrmion ensembles. (a),(c),(e) System with isotropic interactions. (b),(d),(f) System with anisotropic interactions. The simulation temperatures are indicated in each panel. Skyrmions with five and seven neighbors forming lattice defects are colored orange and yellow, respectively. Skyrmions with six neighbors, where the orientational order parameter $|\Psi_{6}(\bm{r}_{i})|$ is close to unity, are colored according to the local lattice orientation $\arg\Psi_{6}(\bm{r}_{i})$ as shown in the colorbar. %skyrmion lattice configurations with isotropic (a), c), e)) and anisotropic (b), d), f) interaction potentials. Colors indicate number of neighbors for each skyrmion according voronoi tesselation. In a) and b), the systems are in a quasicrystalline phase, with dislocation pairs being visible. In c) and d), isolated dislocations appear, with c) being in the hexatic phase. In e) and f), isolated diclinations are visible, the systems are in their respective liquid phase. 
    \caption{Examples of simulations of skyrmion ensembles. (a),(c),(e) System with isotropic interactions. (b),(d),(f) System with anisotropic interactions. The simulation temperatures are indicated in each panel. Skyrmions with five and seven neighbors forming lattice defects are colored orange and yellow, respectively. Skyrmions with six neighbors, where the orientational order parameter $|\Psi_{6}(\bm{r}_{i})|$ is close to unity, are colored according to the local lattice orientation $\arg\Psi_{6}(\bm{r}_{i})$ as shown in the colorbar. %skyrmion lattice configurations with isotropic (a), c), e)) and anisotropic (b), d), f) interaction potentials. Colors indicate number of neighbors for each skyrmion according voronoi tesselation. In a) and b), the systems are in a quasicrystalline phase, with dislocation pairs being visible. In c) and d), isolated dislocations appear, with c) being in the hexatic phase. In e) and f), isolated diclinations are visible, the systems are in their respective liquid phase. 
    }
    \label{fig:Neighbors}
\end{figure}
    
     We performed quasiparticle dynamics simulations following the procedure described in Sec.~\ref{sec:2b} using the two types of interaction potentials determined above. Figure~\ref{fig:Neighbors} shows real-space representations of the simulations at selected temperature values, with the lattice defects highlighted and the phase of the orientational order parameter $\arg\Psi_{6}$ color-coded. The orientation is calculated relative to the horizontal $[1\overline{1}0]$ atomic lattice direction. %horizontal axis. %skyrmion lattices, with isotropic interaction potentials in parts a), c) and e), and anistropic interactions in parts b), d) and f). 
     In panels (a) and (b), we see %lattices
     sections of configurations at relatively low temperatures clearly in the %quasicrystalline
     solid phase, with only non-topological dislocation pairs being visible and $\arg\Psi_{6}$ hardly varying over the simulation area. As we increase the temperature in the %respective 
     systems, we observe the emergence of isolated dislocations in panels (c) and (d), indicative of a hexatic phase according to \ac{KTHNY} theory. While the phase of the orientational order parameter slowly varies over the system for isotropic interactions, for the anisotropic interactions it remains close to $\arg\Psi_{6}=\pi/6$, i.e., the bonds are still formed primarily along the $[2\overline{1}\overline{1}]$ direction. Raising the temperature even further in panels (e) and (f), isolated disclinations appear in both systems, indicating a transition into the liquid phase within \ac{KTHNY} theory. The phase $\arg\Psi_{6}$ varies rapidly for isotropic interactions in panel (e), indicating the loss of orientational order. \\
     Note that the temperatures at which defect unbinding occurs also differ significantly between the two potentials. 
     For the anisotropic interactions in panel (f), it can still be observed that the value of $\arg\Psi_{6}=\psi/6$ is preferred, and the system does not become completely isotropic even at such elevated temperatures. Note that this is different from the observations for anisotropic hard particles in Refs.~\cite{Wojciechowski2004, Anderson2017, Walsh_2016}, where the particles were allowed to rotate leading to isotropy in the liquid phase. In our simulations, the favorable directions for bond formation cannot rotate, because they are locked to the crystallographic axes of the underlying atomic lattice, which remains stable in the investigated temperature range. 
    With that, our work is more closely linked to earlier works discussing phase transitions of particles on periodic substrates~\cite{Nelson_1979}, although the spatially modulated potential created by the substrate is not considered in our simulations. %With that, our work is more closely linked to early work discussing KTHNY \cite{Nelson_1979}, specifically the case of an incommensurate (floating) lattice that breaks continuous rotational symmetry.
      
     %It is to note that the temperatures, at which defect unbinding and therefore a phase transition occurs, differs significantly between the two potentials. 
     While at $T=9~\mathrm{K}$, the system with anisotropic interactions is still in the %quasicrystalline 
     solid phase, the system with isotropic interactions %skyrmions are
     has already melted into an isotropic liquid. This may in part be explained by a slightly larger core size for the anisotropic interaction: note that the hard core of the skyrmions visible in yellow color in Fig.~\ref{fig:Potentials} is hexagonal for anisotropic interactions, while it is circular for isotropic interactions. Since the circle is constructed such a way to touch the sides of the hexagon from the inside, the cores in the case of anisotropic interactions cover a larger area, resulting in a denser packing for the same number of skyrmions. However, the preference for the formation of the bonds along the $[2\overline{1}\overline{1}]$ directions probably plays a more important role in stabilizing the ordered phases. Since each skyrmion admits six neighbors at the six well-defined minima of the anisotropic interaction potential, the formation of lattice defects with a different number of neighbors costs more energy than for the isotropic interactions where the skyrmions can be more easily displaced along the azimuthal direction. %, as the isotropic interaction is a rotation of the smaller \hkl[2-1-1]-direction interaction. It is debatable whether this size difference alone is responsible for the increase in temperature for the unbinding of topological lattice defects. It is more likely that due to the preferred direction of the lattice, the stability thereof increases.

\begin{figure}
    \centering
    \includegraphics[width=0.9\linewidth]{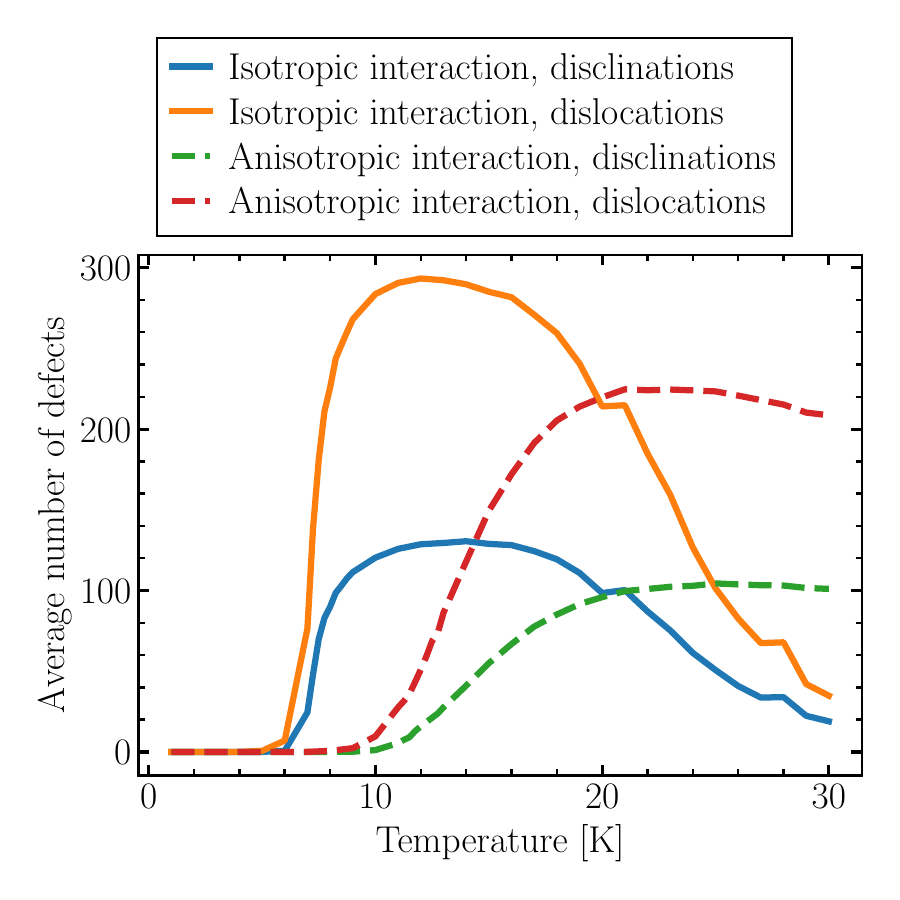}
    \caption{Average number of topologically non-trivial defects %(dislocations and disclinations) 
    as a function of temperature. Solid lines show results for a system with isotropic interactions, dashed lines for a system with anisotropic interactions.}
    \label{fig:Defects}
\end{figure}

    For a more quantitative description of the defects during the melting process using defects, we counted the number of isolated topological lattice defects %throughout the skyrmion lattice 
    and averaged over simulation time for different temperatures, which can be seen in Fig.~\ref{fig:Defects}. Note that we %used a very strict definition of lattice defects, i.e. we do not count defects in large clusters of defects, but 
    strictly only regard dislocations as pairs of skyrmions with five and seven neighbors, and disclinations as skyrmions with a number of neighbors different from six, both of which are surrounded by skyrmions with exactly six neighbors. This excludes other types of defects from the statistics, including dislocation pairs in the %quasicrystalline
    solid phase but also larger clusters of defects emerging at higher temperatures. We find for the isotropic interaction that the first dislocations appear at $T=7~\mathrm{K}$ and the first disclinations at $T=8~\mathrm{K}$. The number of defects increases to a maximum in a very narrow temperature range %quickly 
    for both types. Due to the strict classification of defects, we see that the number of defects starts decreasing %by approximately 
    above $T=12~\mathrm{K}$ after a relatively constant region, which is not due to %to the presence of 
    an increase in order in the %more ordered 
    system, but rather to the emergence of larger clusters that fall outside of our classification. The unbinding of defects according to \ac{KTHNY} theory %The emergence of defects is very fast and 
    indicates a very narrow hexatic phase with dislocations but no disclinations between $T=7~\mathrm{K}$ and $T=8~\mathrm{K}$. For anisotropic interactions, the first dislocations appear only at $T=10~\mathrm{K}$, while disclinations form above $T=12~\mathrm{K}$, restricting the hexatic phase into this temperature regime according to the above classification. The number of defects saturates in a considerably wider temperature range and reaches a lower maximum value than for the isotropic interactions, supporting the argument above that it is more difficult to form topological defects for anisotropic interactions with preferred bond orientations with respect to the atomic lattice.
    
\begin{figure}
    \centering
    \includegraphics[width=0.9\linewidth]{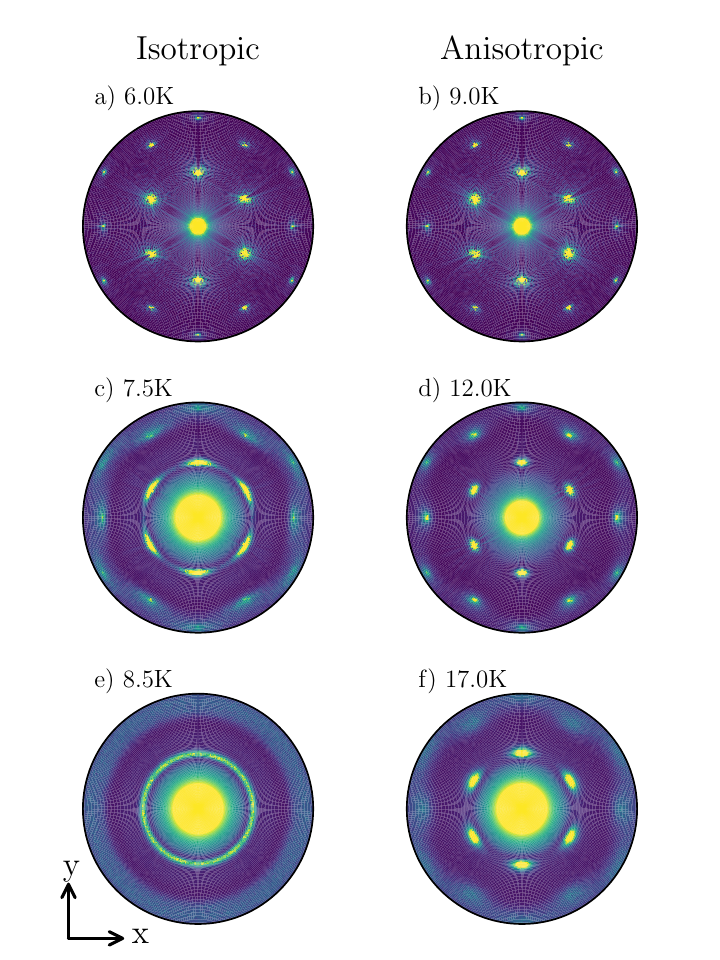}
    \caption{Structure factor $S(\bm{q})$ calculated via Eq. \eqref{eq:structure_factor}. (a),(c),(e) System with isotropic interactions. (b),(d),(f) System with anisotropic interactions. Simulation temperatures are indicated in each panel. %Isotropic interaction shows the known KTHNY-melting: Discrete peaks in a), hexatic ring in c) and isotropic ring in e). The anisotropic interaction shows broadening of the reciprocal lattice vector peaks but never indicates a phase transition.
    }
    \label{fig:SF}
\end{figure}

\begin{figure*}
  \centering
  \includegraphics[width=\textwidth]{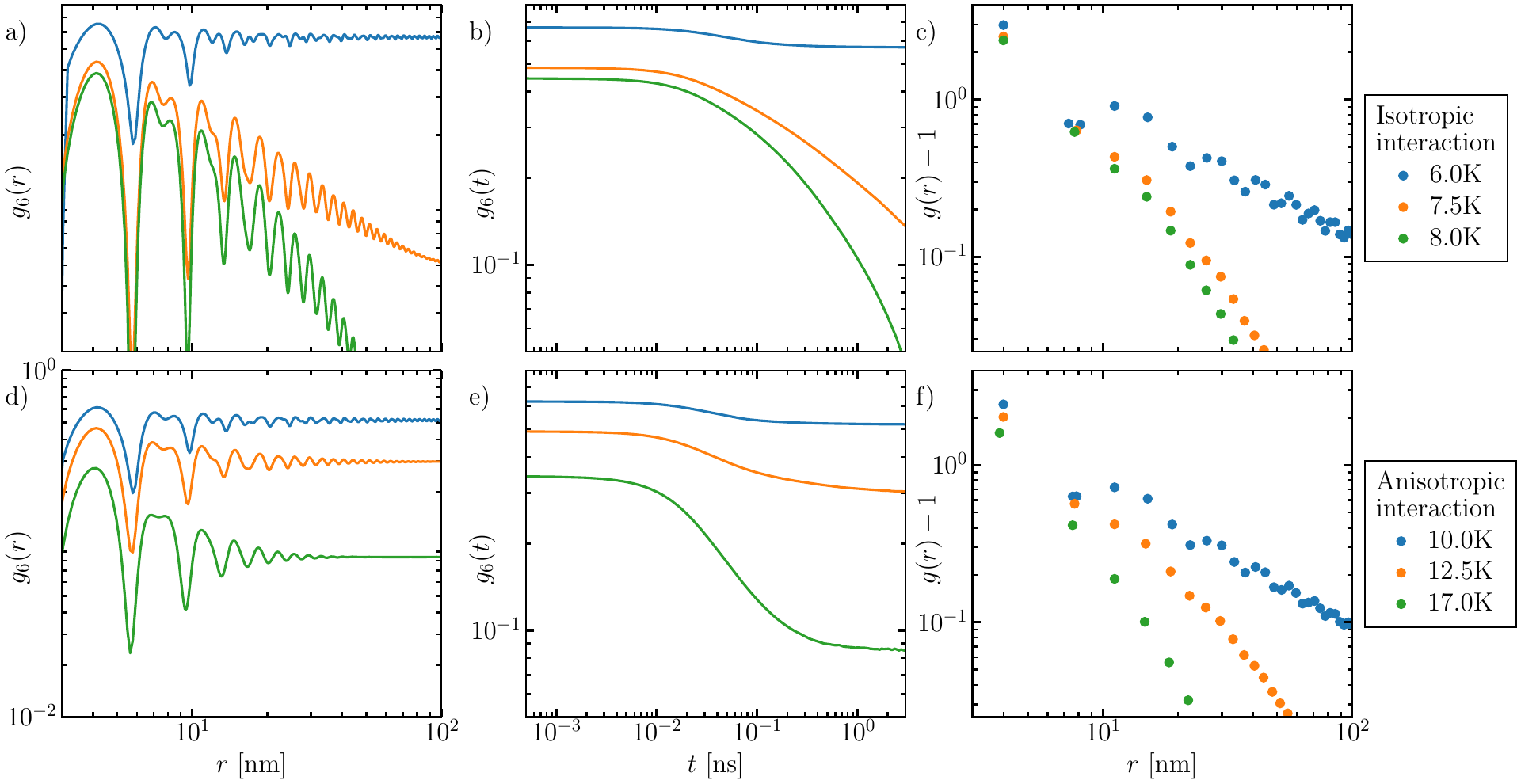}
  \caption{Calculated correlation functions in the system. (a) and (d) Spatial orientational correlation function $g_6(r)$, (b) and (e) temporal orientational correlation function $g_6(t)$, and (c) and (f) positional correlation function $g(r)-1$, %, and mean squared displacement (from left to right), 
  for both (a)-(c) the system with isotropic interactions and (d)-(f) the system with %in panels a)-d), and 
  anisotropic interactions. The functions are plotted on a log-log scale to visually distinguish between algebraic and exponential decays. Only the local maxima of the oscillating function are shown in panels (c) and (f).%in e)-h). For the isotropic interaction, we observe the usual KTHNY melting scenario, with positional order becoming short-ranged at the quasicrystalline to hexatic transition, and orientational order becoming short ranged at the hexatic to isotropic liquid transition. The same behavior is observed in the temporal orientation correlation and the mean squared displacement becoming unbounded. For the anisotropic interaction, we also observe the pair correlation becoming short ranged between $10 \mathrm{K}$ and $12.5 \mathrm{K}$, however the orientation correlation always stays finite in the long range and long time limit.
  }
  \label{fig:wide}
\end{figure*}
    
%     We can also look at the reciprocal lattice structure. For that, we calculate the structure factor 
% \begin{align}
%     S(\bm{k}) = \frac{1}{N} \left\langle \left| \sum_{i,j} e^{- \textrm{i}\bm{k} \cdot (\bm{r}_i-\bm{r}_j)} \right| \right\rangle \label{eq:structure_factor},
% \end{align}
%     with $N$ being the number of particles, $\bm{r}_{i/j}$ being the skyrmion center positions, and $\bm{k}$ being the reciprocal lattice vector. In figure \ref{fig:SF}, we can see the results for isotropic interaction in panels a), c) and e) for increasing temperatures, and in panels b), d) and f) for the anisotropic interaction. For all structure factors, we see large values as $k \rightarrow 0$, which is an artifact due to a finite maximum distance between skyrmion that we consider.
%     \\
        We start the analysis of the correlation functions with the structure factor from Eq.~\eqref{eq:structure_factor}, shown in Fig.~\ref{fig:SF}.
At low temperature, we identify sharp peaks at the reciprocal-lattice vectors of the %quasicrystalline 
    solid phase both for isotropic (Fig.~\ref{fig:SF}(a)) and anisotropic (Fig.~\ref{fig:SF}(b)) interactions. With increasing temperature in panels (c) and (d), the peaks widen along the azimuthal direction, and also along the radial direction primarily for peaks further away from the center. This hints at the loss of quasi-long-ranged positional order, while orientational order is preserved to an extent, which is consistent with a hexatic phase~\cite{Marcus1996}. The most pronounced difference between the different types of interactions is shown in Figs.~\ref{fig:SF}(e) and (f): for isotropic interactions, rings with uniform intensity can be observed around the center, which signals a transition into the isotropic liquid phase. However, the sixfold symmetry of the structure factor is preserved for anisotropic interactions up to $T=17$~K, where there is already a very high number of dislocations and disclinations in the system, cf. Fig.~\ref{fig:Defects}. This high-temperature liquid phase still shows signatures of the preferred bond orientations induced by the atomic lattice.
    % We find the usual KTHNY behavior for the isotropic interaction, with discrete peaks showing sixfold symmetry in the qusicrystalline phase, which with increasing temperature becomes a ring, still showing six equidistant maxima around the center. With a further increase in temperature, the ring becomes isotropic, which signals a transition into a the isotropic liquid phase \cite{Marcus1996}. We see the usual KTHNY behavior in this situation. \\
    % Considering the anisotropic interaction, the behavior differs vastly: We observe the usual structure for the quasicrystalline phase, but with increasing temperature, the peaks never transition into an anisotropic ring. Instead, we observe discrete peaks with sixfold symmetry, even tough we see a widening with increasing temperature. \\
    % From these findings, we could interpret that the transition into the liquid phase takes place at higher temperatures than we considered. However, previously we found that at the same temperatures we considered, the lattice already shows many topological lattice defects, which indicate that we are indeed in a liquid at a temperature of $17 \mathrm{K}$. Instead we see a preferred global lattice rotation, which suppresses phase transition indicators in reciprocal space. 

    In Fig.~\ref{fig:wide}, the temperature dependence of 
    %To better understand the phase transition, we calculate the rotational and positional correlation functions. In addition to 
    the spatial orientational correlation function from Eq.~\eqref{eq:g6r}, the temporal orientational correlation function from Eq.~\eqref{eq:g6t}, and the maxima of the positional correlation function from Eq.~\eqref{eq:pairCorrelation} is shown for the two types of interaction potentials. 
    %We averaged all skyrmion trajectories $t=0$ staring points. The results of these four calculations can be seen in Figure \ref{fig:wide}, where panels a)-c) show quantities for isotropic interaction and panels d)-f) for anisotropic interaction. \\
    At the lowest temperature for both the isotropic ($T=6$~K) and the anisotropic ($T=10$~K) interactions, the spatial and temporal orientational correlation functions converge to a finite value, while the positional correlation function decays algebraically, identifiable as a linear decrease in the log-log plots. These signatures are indicative of the %quasicrystalline 
    solid phase.
    %We first compare the low temperature case, which is $6 \mathrm{K}$ for the isotropic interaction and $10 \mathrm{K}$ for the anisotropic case. We can see that for both potentials, the spatial orientation correlation function $g_6(r)$ does not go to zero (panels a) and d)), nor does the temporal orientational correlation $g_6(t)$, as seen in panels b) and e). This is indicative of the quasicrystaslline phase. The pair correlation function $g(r)$ shows algebraic decay in panels c) and f). The mean squared displacement increases very slowly at high times $\Delta t$, also indicative of the quasicrystalline phase.  \\
    At $T=7.5$~K for the system with isotropic interactions, both $g_6(r)$ and $g_6(t)$ decay algebraically, while the positional correlation function decays exponentially, which is faster than the linear decay in the log-log plot. %{\red{I do not see $g_6(t)$ decaying exponentially as was written before.}} {\blue{It's just faster than algebraically, maybe we should reframe it that way}}
    This indicates a quasi-long-ranged orientational and short-ranged positional order, consistent with a hexatic phase. Contrary to this scenario, for anisotropic interactions the orientational correlation function converges to a finite value both at $T=12.5$~K and at $T=17$~K, with this order parameter decreasing with increasing temperature. The positional order is also short ranged for the anisotropic interactions. %The persistent orientational order together with the short-ranged positional order resembles a hexatic %or quasicrystalline solid 
    %phase. 
    In contrast, the orientational correlations also becomes short ranged at $T=8$~K for the isotropic interactions, indicating a transition into the isotropic liquid phase in agreement with the conclusions drawn from the unbinding of defects.
    % As we increase the temperature, both systems start to differ, with the orientational correlation function $g_6(r)$ decaying algebraically. For $g_6(t)$ we also observe decaying behavior, which even may be interpreted as exponential. This is not entirely consistent with the hexatic phase, however, since we are close to the phase transition into the isotropic liquid phase, this behavior may follow. Contrary to this scenario, we do not see decay to zero in the anisotropic case, rather the limits $\lim\limits_{r\rightarrow \infty} g_6(r)$ and $\lim\limits_{t \rightarrow \infty} g_6(t)$ get lowered compared to the lower temperatures. As for the pair correlation function, we see large deviations from the quasicrystalline case, with it decaying very quickly in the isotropic case, and even tough slightly slower in the anisotropic case, still much faster than before with a non algebraic decay. This indicates a short ranged translational order. For the isotropic case this is consistent with the KTHNY melting scenario, especially as we further increase the temperature, we see orientation correlation becoming short ranged, as well as non-algebraic time decay. \\
    % In the anisotropic interaction, we observe that neither $g_6(r)$ nor $g_6(t)$ approach zero, but rather some constant value even at the higher temperatures, where the translational correlation is already short-ranged. This is not consistent with the KTHNY melting scenario and shows the emergence of a phase with global orientational order and short-ranged translational order, i.e. an anisotropic liquid phase. 

\begin{figure}
    \centering
    \includegraphics[width=0.9\linewidth]{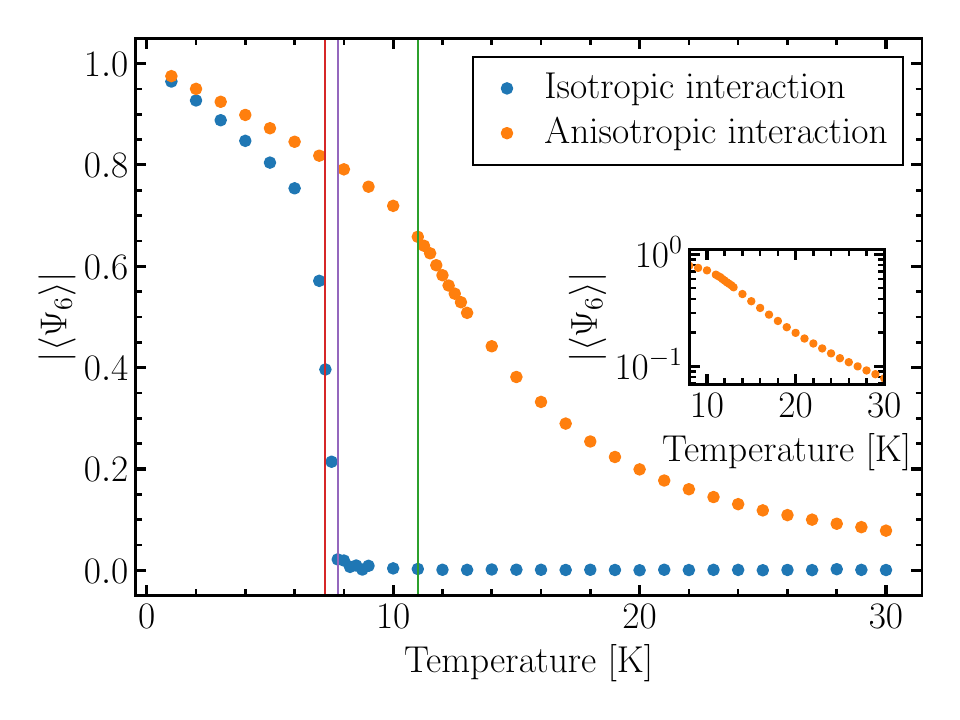}
    \caption{Average bond order parameter $|\langle  \Psi_6  \rangle|$. Vertical lines show phase transitions determined from the decay of orientational and positional correlation functions. Red line denotes solid to hexatic, purple line hexatic to liquid transition for isotropic interaction ($T\approx7~\mathrm{K}$ and $T\approx8~\mathrm{K}$ respectively). Green line denotes solid to anisotropic liquid transition for anisotropic interaction ($T\approx11~\mathrm{K}$). Inset shows the values for the anisotropic interactions on a logarithmic scale to visualize the exponential decay.}
    \label{fig:g6inf}
\end{figure}

We calculated the average bond order parameter $| \langle \Psi_6 \rangle| $, as shown in Fig.~\ref{fig:g6inf}. While in the system with isotropic interactions, this order parameter reaches zero at around $T=8$~K as the system transforms into the isotropic liquid phase, in the system with anisotropic interactions the orientational order persists at all investigated temperatures up to 30~K. The exponential decay as seen in the inset suggests a persistence to even higher temperatures within the scope of validity for our model.
%One may now ask, whether or not there is a transition into an isotropic liquid phase. To answer this, we calculated the long-ranged limit of $g_6(r)$ for various temperatures and show this in Figure \ref{fig:g6inf}. While for the isotropic case, the curve behaves as expected, starting at finite values and then quickly going to zero, this is not the case for the anisotropic interaction, where this decay is more gradual, and we seee an exponential decay towards zero, but not quite reaching it. We therefore conclude that in our system, we never reach a fully isotropic liquid state with anisotropic skyrmion-skyrmion interactions. 

Overall, for the isotropic interactions the correlation functions indicate two phase transitions: first from a solid phase with quasi-long-ranged positional and long-ranged orientational order to a hexatic phase with short-ranged positional and quasi-long-ranged orientational order at $T\approx7~\mathrm{K}$, then from the hexatic to the isotropic liquid phase where all correlations are short ranged at $T\approx8~\mathrm{K}$. The change in the decay of the correlation functions agrees with the onset of first the dislocations, then the disclinations in Fig.~\ref{fig:Neighbors}. These observations are consistent with KTHNY theory. For the anisotropic interactions, we find that the anisotropy induced by the atomic lattice enforces orientational order in the system at all investigated temperatures. The positional correlation function changes from an algebraic to an exponential decay at $T\approx11~\mathrm{K}$. This agrees with an inflection point in the orientational order parameter in Fig.~\ref{fig:g6inf}, above which it decays exponentially. We find no other signature of a phase transition, and identify the low-temperature phase as a solid and the high-temperature phase a liquid with anisotropic skyrmion orientations induced by the interaction potential. Although the hexatic phase is also characterized by quasi-long-ranged orientational correlations, in the anisotropic case the orientational order is not caused by spontaneous symmetry breaking but by the atomic lattice. Therefore, the onset of dislocations and disclinations in Fig.~\ref{fig:Defects} at different temperatures cannot be connected to two separate phase transitions, either.

\subsection{Machine-learning classification of phases}

We also used a machine-learning algorithm to identify possible phases during the melting transition. For %a machine learning characterization
this purpose, we rasterized the local bond-order parameter $\Psi_{6,i}$. We created a grid and averaged the complex argument $\text{arg}(\Psi_{6,i})$ and the absolute value $|\Psi_{6,i}|$ for all skyrmions in each cell of the grid. Using periodic boundary conditions, each rasterization was then also shifted by a random amount in the $x$ and $y$ coordinates. This was done to minimize overfitting to a particular configuration of the skyrmion lattice. %Using configurations for different simulation times, we are able to create well separated clusters for rasterization ranging from $4\times4$ to $32\times32$. 
%Figure~\ref{fig:PU} shows the phase diagram for a 
We used a rasterization grid of $40\times40$, where each cell had on average about 10.4 skyrmions in them. For finer grids there was a chance of having no skyrmions in a cell, and coarser grids produced less separated clusters. The averaged local bond-order parameter was then used as an input for the k-means clustering algorithm~\cite{mcqueen1967some} with the implementation from the Python package scikit-learn~\cite{scikit-learn} for identifying possible phases. %is used. %A clear and unambiguous identification of the hexatic phase was not possible with this approach.

\begin{figure}
    \centering
    \includegraphics[width=1\linewidth]{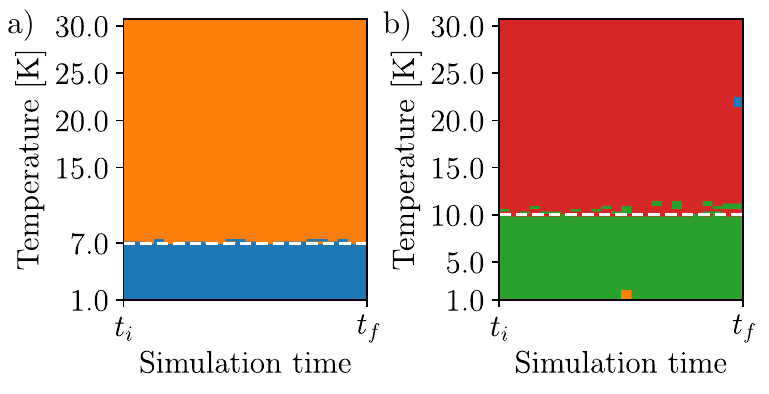}
    \caption{Phase diagrams obtained from the machine-learning algorithm. %determined phases. 
    Systems with isotropic and anisotropic interactions are shown in panels (a) and (b), respectively. Colors indicate the four different clusters. Along the horizontal axis different snapshots are shown from a simulation performed at a fixed temperature between initial time $t_i$ and final time $t_f$. % a) and b) show the phase diagrams of the isotropic and the anisotropic case, respectively. The $y$ axis is the temperature and the $x$ axis is the simulation time. $t_i$ stands for the initial time step and $t_f$ for the final one. 
    White dashed lines are guides to the eye identifying %roughly mark 
    the critical temperatures of the predicted phase transitions.
    }
    \label{fig:PU}
\end{figure}

The phase diagrams deduced from the machine-learning algorithm are shown in Fig.~\ref{fig:PU}. The algorithm identified a low-temperature and a high-temperature phase for both systems with isotropic and anisotropic interactions, respectively. Although the two datasets for different interactions were combined during the analysis, the algorithm managed to separate the two types of interactions, apart from a few outlying configurations. This is surprising in the low-temperature %quasicrystalline 
solid phases, where our previous analysis could not identify a clear difference. However, in the high-temperature limit the isotropic liquid phase and the liquid phase with enforced orientational order were also separated by the algorithm. The two clusters in each figure are separated by an almost horizontal line into a low-temperature and a high-temperature regime. %that different phases may be observed at different temperatures, but different configurations at the same temperature belong to the same phase far from the transition. 
Close to the transition line, configurations obtained from the same simulation may belong to different clusters; this may either be an indication for phase coexistence or designate data points located between two well-defined clusters. Note that the machine-learning algorithm only identified a single phase transition in both systems, correlating with the inflection point in the orientational order parameter in Fig.~\ref{fig:g6inf}. %, which is close to the transition from the %quasicrystalline 
%solid to the hexatic phase in both systems as deduced from the correlation functions. 
It is possible that the algorithm could not separate the hexatic phase from the isotropic liquid phase for isotropic interactions because the former is found only in a narrow temperature range, which may contain an insufficient number of data points to be possible to be identified as a separate cluster. For the system with anisotropic interactions, the temperature of the single identified phase transition is close to the value determined from the analysis of the correlation functions above. %the order parameter varies smoothly in the high-temperature regime, and it is not clear that two separate phases should exist in this region. Due to the averaging over the skyrmions inside the discretization cells, the machine-learning algorithm most likely did not obtain information about the topological defects in the system, which were also used to characterize the phases above.

\section{Conclusions}

We studied the influence of the anisotropy of the skyrmion-skyrmion interaction potential on the melting of the skyrmion lattice. We determined a spatially anisotropic interaction potential for skyrmions in the (Pt$_{0.95}$Ir$_{0.05}$)/Fe/Pd(111) system from atomistic simulations, which prefers bond formation along the $[2\overline{1}\overline{1}]$ and symmetrically equivalent directions. For comparison, we also constructed an isotropic potential with the same depth of the potential minima. We performed dynamical simulations treating the skyrmions as quasiparticles to study the melting transition. By analyzing the topological defects in the system, we found an onset of dislocations at $T=7$~K and at $T=10$~K, and the appearance of disclinations at $T=8$~K and at $T=12$~K for systems with isotropic and anisotropic skyrmion-skyrmion interactions, respectively. We concluded that the preferred directions of bond formation in the anisotropic potential hinder the formation of defects, thereby increasing the stability of the ordered phase. From the analysis of the correlation functions, we found that quasi-long-range positional correlations persist in the system approximately up to the temperature where the dislocations appear for both types of interactions, identifying this low-temperature phase as a solid state. %in qualitative agreement with the \ac{KTHNY} theory. 
For the isotropic interactions, the orientational order remains quasi-long-ranged up to the onset of disclinations, making it possible to identify a second transition at around $T=8$~K from a hexatic to an isotropic liquid phase. The observations for this system are in qualitative agreement with \ac{KTHNY} theory. In contrast, we found that in the case of anisotropic interactions the orientational order parameter remains finite, only decreasing asymptotically with temperature. While quasi-long-ranged orientational order is characteristic of a hexatic phase instead of a liquid phase, here the order is enforced by the anisotropy of the interactions; therefore, we identified the high-temperature phase as a liquid with induced anisotropy for this system. %, which is characteristic of a hexatic phase but cannot be reconciled with an isotropic liquid phase. 
We also characterized the obtained local orientational order parameters in the system using a machine-learning algorithm, which successfully distinguished between the two types of interactions and between solid and non-solid phases, but could not identify the narrow temperature range where the hexatic phase was found previously for isotropic interactions.

Our simulations for skyrmions with short-range attraction found indications for a hexatic phase, similarly to previous experimental studies for skyrmions with purely repulsive interactions~\cite{Huang2020,Gruber2025}. Further investigations of the system with isotropic interactions may be possible to elucidate the first-order or second-order nature of the phase transitions, which could clarify whether \ac{KTHNY} theory or a competing interpretation is best applicable in this scenario. However, this will require increasing the number of particles in the simulations further, because the correlation lengths often reach very high values in similar simulations~\cite{Anderson2017}. 
Further research could investigate tuning the skyrmion-skyrmion interaction potentials to widen the hexatic phase. This could include the computation of density-temperature phase diagrams. %Dynamical parameters, such as gyrocoupling $\mathcal{G}$ or Gilbert damping $\alpha$ are expected to not modify phase formation. 

Our simulations take the anisotropy in the interactions as a mechanism of breaking continuous rotational symmetry into account, and therefore differ from works with non-circular particle shapes which are allowed to rotate~\cite{Wojciechowski2004, Anderson2017, Walsh_2016}. Instead, our work is more closely linked to particles within an periodic potential induced by the lattice. %, in the incommensurate limit.

Previous simulations for repulsive skyrmions with isotropic interactions in Ref.~\cite{Nishikawa2019} found a direct transition from the solid to the isotropic liquid phase, which was attributed to the strong periodic potential exerted by the underlying atomic square lattice on the small-scale skyrmions. While this periodic potential was not considered in our quasiparticle simulations, the influence of the atomic lattice was taken into account in the anisotropy of the interactions. This also resulted in a direct solid-to-liquid transition, but sixfold orientational order also persisted in the liquid phase, in contrast to the isotropic liquid phase identified in Ref.~\cite{Nishikawa2019}.

Systems where skyrmions are stabilized by the competition between ferromagnetic and antiferromagnetic magnetic exchange interactions, such as the (Pt$_{0.95}$Ir$_{0.05}$)/Fe/Pd(111) system studied here, offer further prospects for studying the role of the anisotropy on the phase formation of quasiparticles. For example, such systems allow for the stabilization of different types of particle-like spin configurations~\cite{Leonov2015,Rozsa2017}. These include antiskyrmions, which besides having anisotropic interaction potentials, may also rotate on the atomic lattice~\cite{Weissenhofer2020}, introducing a further degree of freedom which may influence the melting transition. The investigations could further be extended to ensembles containing both skyrmions and antiskyrmions, which have been demonstrated to prefer a non-triangular arrangement already in the solid phase~\cite{Leonov2015}.

\begin{acknowledgments}
We would like to thank Markus Wei{\ss}enhofer and Elena Y. Vedmedenko for fruitful discussions. We acknowledge funding by the Deutsche Forschungsgemeinschaft (DFG, German Research Foundation) via SFB 1432 and Project Nos. 403502522 and 514141286, by the National Research, Development, and Innovation Office (NRDI) of Hungary under Project Nos. FK142601 and ADVANCED 149745, by the Ministry of Culture and Innovation and the National Research, Development and Innovation Office within the Quantum Information National Laboratory of Hungary (Grant No. 2022-2.1.1-NL-2022-00004), and by the Hungarian Academy of Sciences via a J\'{a}nos Bolyai Research Grant (Grant No. BO/00178/23/11). We are grateful for the computational support and resources provided by the Scientific Compute Cluster of Univerity of Konstanz (SCCKN) for the simulations of the skyrmion lattices.
\end{acknowledgments}

\bibliographystyle{apsrev4-2}
\bibliography{bibfile}

\end{document}